\documentclass[aps,pra,twocolumn,superscriptaddress,10pt,nofootinbib,longbibliography,floatfix,article]{revtex4-2}
\usepackage{ORI_Group_style, symbols}
\usepackage[english]{babel}
\usepackage{subfigure}

\hypersetup{colorlinks,linkcolor=black,citecolor=blue,urlcolor=blue}

\begin{document}

\title{Large Quantum Delocalization of a Levitated Nanoparticle using Optimal Control: Applications for Force Sensing and Entangling via Weak Forces}

\author{T. Weiss}
\affiliation{Institute for Quantum Optics and Quantum Information of the Austrian Academy of Sciences, 6020 Innsbruck, Austria}
\affiliation{Institute for Theoretical Physics, University of Innsbruck, 6020 Innsbruck, Austria}

\author{M. Roda-Llordes}
\affiliation{Institute for Quantum Optics and Quantum Information of the Austrian Academy of Sciences, 6020 Innsbruck, Austria}
\affiliation{Institute for Theoretical Physics, University of Innsbruck, 6020 Innsbruck, Austria}

\author{E. Torrontegui}
\affiliation{Departamento de F\'{\i}sica, Universidad Carlos III de Madrid, 28911 Legan\'es (Madrid), Spain}
\affiliation{Instituto de F\'{\i}sica Fundamental IFF-CSIC, Calle Serrano 113b, 28006 Madrid, Spain}

\author{M. Aspelmeyer}
\affiliation{Vienna Center for Quantum Science and Technology, Faculty of Physics, University of Vienna, A-1090 Vienna, Austria}
\affiliation{Institute for Quantum Optics and Quantum Information, Austrian Academy of Sciences, A-1090 Vienna, Austria}

\author{O. Romero-Isart}
\affiliation{Institute for Quantum Optics and Quantum Information of the Austrian Academy of Sciences, 6020 Innsbruck, Austria}
\affiliation{Institute for Theoretical Physics, University of Innsbruck, 6020 Innsbruck, Austria}

\date{\today}

\begin{abstract}

We propose to optimally control the harmonic potential of a levitated nanoparticle to quantum delocalize its center-of-mass motional state to a length scale orders of magnitude larger than the quantum zero-point motion. Using a bang-bang control of the harmonic potential, including the possibility to invert it, the initial ground-state-cooled levitated nanoparticle coherently expands to large scales and then contracts to the initial state in a time-optimal way. We show that this fast loop protocol can be used to enhance force sensing as well as to dramatically boost the entangling rate of two weakly interacting nanoparticles. We parameterize the performance of the protocol, and therefore the macroscopic quantum regime that could be explored, as a function of displacement and frequency noise in the nanoparticle's center-of-mass motion. This noise analysis accounts for the sources of decoherence relevant to current experiments. 

\end{abstract}

\maketitle

\begin{figure}
    \centering
    \includegraphics[width=\columnwidth]{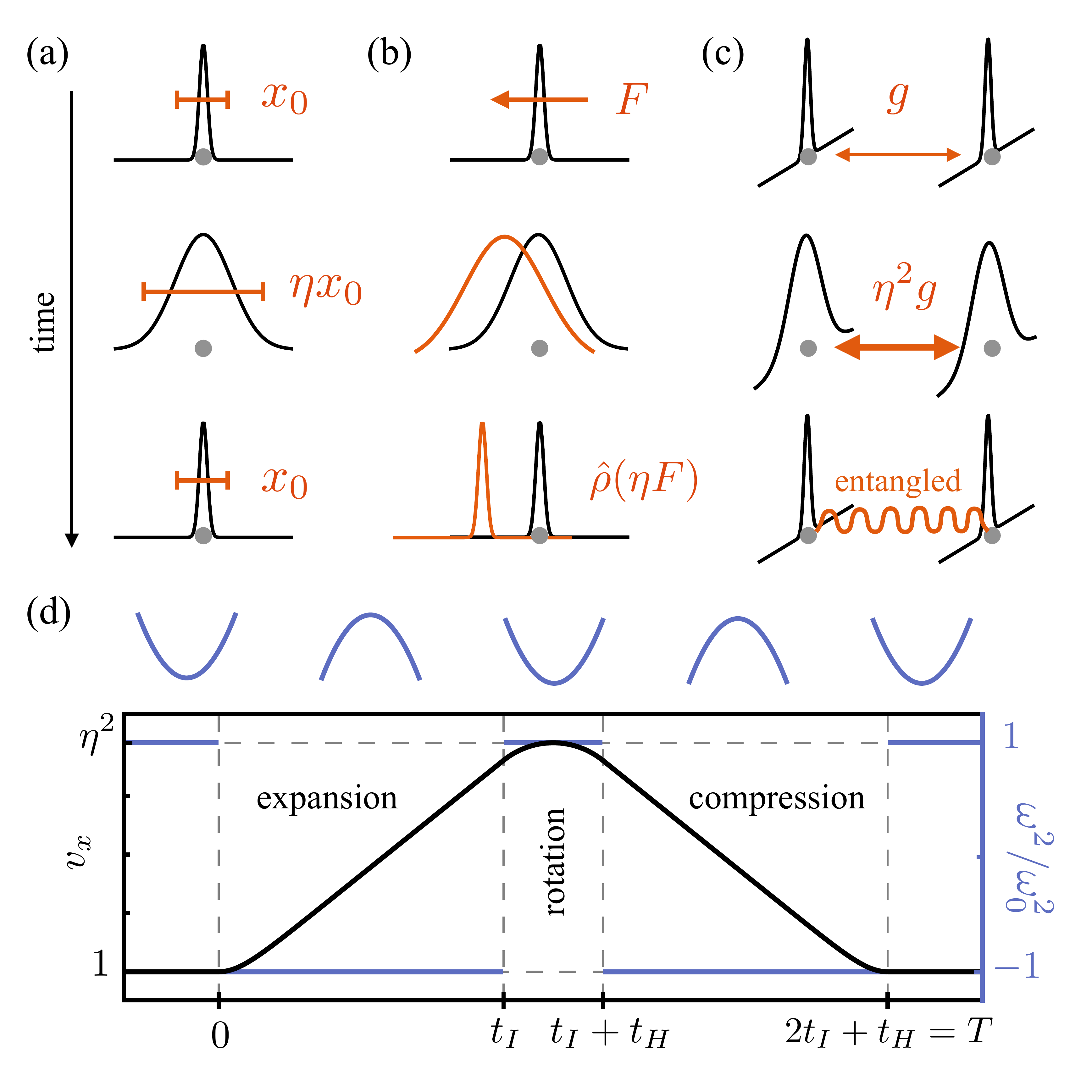}
    \caption{Applications of the proposed loop protocol: (a) coherent expansion to large scales, (b) static force sensing, and (c) entangling two particles via weak forces. (d) Dimensionless variance $\varianceX \equiv \avg{\position^2}/\xzpf^2$ (solid black line, left axis in logarithmic scale) and $\freqg^2/\freq^2$ of the harmonic potential (solid blue line, right axis in linear scale) as a function of time in the optimal bang-bang loop.   }
    \label{fig:1}
\end{figure}

A levitated nanoparticle in high vacuum is a promising system to explore quantum mechanics at large scales because of three reasons. (i) The center-of-mass motion of a nanoparticle of mass $\mass$ in the regime of $10^8$ to $10^{11}$ atomic mass units (AMUs) can be prepared in a pure quantum state via ground-state cooling in a tight harmonic potential of frequency $\omega_0$ ($\sim 2\pi \times 100$ kHz)~\cite{Romero_Isart_SuperposLiving_2010, Chang_levitatedNano_2010,PhysRevA.81.023826,PhysRevA.83.013803,Delic_gsCooling_2020,magrini_optimal_2020,tebbenjohanns_quantum_2021}.
The corresponding zero-point motion $\xzpf \equiv \sqrt{\hbar/(2 \mass \freq)}$ is minute ($\sim 10^{-12}$~m). (ii) The harmonic potential can be tuned to induce dynamics in which the position probability distribution is expanded to scales larger than $x_0$, ideally approaching the length scale given by the size of the nanoparticle ($\sim 10^{-7}$~m)~\cite{PhysRevLett.107.020405,PhysRevA.84.052121,bateman_near-field_2014,Hebestreit_staticForceSensing_2018}. (iii) Levitation in ultra-high vacuum provides a high degree of isolation of its center-of-mass motion, enabling the induced dynamics to be coherent. In this Letter, we combine these three ingredients in an {\em optimal} way to propose feasible nanomechanical experiments~\cite{Aspelmeyer_OptomechanicsReview_2014} exploring and exploiting quantum mechanics at scales orders of magnitude larger than the zero-point motion, see \figref{fig:1}. 
This proposal is particularly timely given the recent experiments on cooling an optically levitated nanoparticle into the quantum regime~\cite{Delic_gsCooling_2020,PhysRevLett.124.013603,Kamba2020,magrini_optimal_2020,tebbenjohanns_quantum_2021}. 
In contrast to other significantly more challenging proposals~\cite{PhysRevLett.107.020405,bateman_near-field_2014,PhysRevD.92.062002,Wan_MacroscopicSuperpositionForce_2016,Pino_2018}, here we do not require to prepare quantum superposition states.

More specifically, we propose to dynamically manipulate the harmonic potential of a nanoparticle, including the possibility to invert it, to first coherently expand its state to large scales and, then, compress it back to its initial state in a time-optimal way. By measuring the purity of the final state (\eg~via its center-of-mass temperature) as well as the position probability distribution at the point of maximum expansion (on different experimental runs), one could conclude that the center-of-mass expanded coherently to scales larger than the zero-point motion~\figref{fig:1}(a). In addition, returning to the initial contracted state is beneficial both to minimize decoherence, as the expanded state is very fragile, and to facilitate the required repetition of experimental runs. The loop protocol can be used to  explore quantum physics of massive objects at large scales, thereby falsifying collapse models~\cite{PhysRevA.84.052121,RevModPhys.85.471}. The loop protocol can also be used for force sensing~\cite{Hebestreit_staticForceSensing_2018}, as the presence of a static force critically alters the final state of the protocol, see \figref{fig:1}(b). Furthermore, if two nanoparticles interact via a weak force (\eg~Coulomb~\cite{Frimmer_ChargeNanoparticle_2017}, Casimir~\cite{Emig_CasimirForces_2007}, gravity~\cite{Marletto_GravitationalEntanglement_2017, Bose_GravitationalEntanglement_2017}), the loop protocol dramatically enhances the entangling rate due to the expansion of the wave function, see \figref{fig:1}(c). In the following, we describe and analyze these ideas in detail by taking into account the effect of displacement and frequency noise in the nanoparticle, which encompass the relevant sources of decoherence in current experiments with levitated nanoparticles~\cite{Carlos_CoolingTheory_2019,PhotonRecoil_Jain_Lukas_2016}.

Let us consider the center-of-mass motion of a nanoparticle along the $x$-axis. The motion along the other axes as well as rotational degrees of freedom are assumed to be either decoupled or to effectively induce displacement and frequency noise. We assume the experimental possibility to dynamically control a harmonic potential in the form $V(\positionC, t) = \mass\freqg^2(t)\positionC^2/2$ with $\freqg^2(t) \in [- \freq^2,\freq^2]$. That is, the harmonic frequency that determines the strength of the spring constant is upper bounded to $\freq$, but the potential can be inverted. We are interested in the optimal control problem to transition between states in a time-optimal way~\cite{salamon_maximum_2009} in order to be faster than decoherence. With potentials of the form $V(\positionC, t)$,  the so-called bang-bang solutions are known to be time optimal~\cite{stefanatos_frictionless_2010,torrontegui_sta_2013}. Bang-bang solutions make $\freqg^2(t)$ alternate, with sudden changes, between the two extreme available values $\pm \freq^2$. Hereafter we thus consider the two Hamiltonians given by
\begin{equation}
   \Hplusminus = \frac{\momentum^2}{2\mass} \pm \frac{1}{2}\mass\freq^2\position^2=\frac{\hbar\freq}{4}\left(\unitlessMomentum^2 \pm \unitlessPosition^2 \right).
   \label{eq:Hamiltonian}
\end{equation}
Here $\position=\unitlessPosition \xzpf$ and $\momentum=\unitlessMomentum\pzpf$, with $\coms{\position}{\momentum}=\im \hbar$, are the position and momentum operators. We have defined dimensionless operators using the zero-point motion position $\xzpf$ and momentum $\pzpf \equiv \hbar/(2 \xzpf)$. We define the thermal state of $\Hplus$ as $\densityMatrixThermal = \sum_{n=0}^\infty P(n) \ket{n} \bra{n}$, where $P(n) = [\bar n/(1+\bar n)]^n (1+\bar n)^{-1}$, $\ket{n}$ with $n=0,1,\ldots$ is a Fock state ($\Hplus \ket{n} =\hbar \freq (n+1/2) \ket{n}$) and $\bar n$ the phonon mean number.

The loop protocol proposed in this Letter, see \figref{fig:1}(d), is based on the following key equality
\be \label{eq:bangbang}
e^{-\im \Hinv \timeInverted/\hbar} e^{-\im \Hplus \timeHarmonic/\hbar}e^{-\im \Hinv \timeInverted/\hbar} = e^{-\im \Hplus \timeHarmonic/\hbar},
\ee 
which holds for $\timeHarmonic  \freq = \pi (2 l +1) / 2$ with $l=0,1,\ldots$ and arbitrary $\timeInverted$, see \cite{SupplementalMaterial}. The left hand side of \eqnref{eq:bangbang} is a unitary operator, labeled by $\Uloop$, that describes the time evolution of the loop protocol from $t=0$ to $t=T \equiv  2 \timeInverted + \timeHarmonic$ in three steps: (1)~inverted harmonic potential $\Hinv$ for $0 < t < \timeInverted$, (2)~harmonic potential $\Hplus$ for $ \timeInverted< t <  \timeInverted + \timeHarmonic$ and (3)~inverted harmonic potential for  $\timeInverted + \timeHarmonic < t <  2\timeInverted + \timeHarmonic$. The loop protocol starts and ends with the particle in the harmonic potential, that is, one has $\Hplus$ for $t<0$ and $t>\totalTime$.  Equation \eqref{eq:bangbang} shows that an initial state $\densityMatrix(0)$ evolves during the loop protocol into $ \densityMatrix(\totalTime)= \Uloop \densityMatrix(0) \Udloop = e^{-\im \Hplus \timeHarmonic/\hbar} \densityMatrix(0)  e^{\im \Hplus \timeHarmonic/\hbar}$. The state at $t=\totalTime$ is thus very similar to the one at $t=0$. In fact, $\densityMatrix(\totalTime) = \densityMatrix(0)$ whenever $\densityMatrix(0)$ is diagonal in the Fock basis (\eg~thermal state). However, the state at $t=\totalTime/2$ is dramatically different as it has expended and squeezed due to the action of $\Hinv$. To quantify the expansion, we define the coefficient $\eta \equiv \stdDev (T/2) / \stdDev(0)$ where  $\stdDev(t) \equiv \spares{\expect{\unitlessPosition^2(t)} - \expect{\unitlessPosition(t)}^2}^{1/2} $. If the initial state is thermal $\densityMatrix(0)=\densityMatrixThermal$, the expansion coefficient is given by $\eta = \exp (\timeInverted \freq)$, that is, it exponentially grows with the time invested in the inverted potential~\cite{Romero_Isart_CoherentInflation_2017}.
The motional quantum state at $t=T/2$ is a highly-squeezed state with large spatial extension, which is very different from a large coherent state with small spatial extent that has been realised with an ion using a bang-bang protocol where the center of the trap is displaced~\cite{alonso_generation_2016}.

Let us discuss how we model noise and decoherence in the loop protocol. We remark that during the protocol $0<t<T$, there is neither active nor passive cooling acting on the nanoparticle. Hence, we consider displacement ($\indexnoise=1$) and frequency  ($\indexnoise=2$) noise, as described by the master equation
\begin{equation} \label{eq:ME}
\partial_t \densityMatrix = \frac{1}{\im \hbar} \com{ \Hplusminus}{\densityMatrix} -\decoh_1\coms{\unitlessPosition}{\coms{\unitlessPosition}{\densityMatrix}}  -\decoh_2\coms{\unitlessPosition^2}{\coms{\unitlessPosition^2}{\densityMatrix}}.
\end{equation} 
This master equation is the result of averaging the noise~\cite{Schneider_decoherenceIonTraps_1999} described by the stochastic Hamiltonian $\Hplusminus + \Hnoise (t)$, with
\begin{equation}
\Hnoise (t) = \frac{\hbar \freq}{4} \spare{ 2 \fone(t) \unitlessPosition + \ftwo(t) \unitlessPosition^2}.
\end{equation}
Here, $\unitlessForce_\indexnoise(t)$ is a stochastic variable with zero mean and power spectral density given by $\PSD_\indexnoise (\freqg) \equiv (2 \pi)^{-1} \intall \text{d} \tau  \avg{\unitlessForce_\indexnoise(t) \unitlessForce_\indexnoise(t+\tau)} e^{\im \freqg \tau} $ (here $\avg{\cdot}$ is an ensemble average over the stochastic variable). In \eqnref{eq:ME}, the decoherence rates are then given by $\decoh_\indexnoise \equiv \pi \freq^2 \PSD_\indexnoise (\indexnoise \freq) / 4^\indexnoise$. Equation \eqref{eq:ME} can also be derived by tracing out an Ohmic bath modelled as a set of quantum harmonic oscillators in the high-temperature limit~\cite{breuer_theory_2010}, see~\cite{SupplementalMaterial}. Standard sources of decoherence in experiments with levitated nanoparticles (\eg~laser recoil heating, blackbody radiation, vibrations, intensity field fluctuations) can be modelled using \eqnref{eq:ME} with the corresponding contribution to the decoherence rate $\decoh_\indexnoise$~\cite{Carlos_CoolingTheory_2019,PhotonRecoil_Jain_Lukas_2016}. In addition, note that fluctuations in the switching times are already taken into account as they are equivalent to  frequency noise. This is explicitly shown in~\cite{SupplementalMaterial}. Decoherence due to scattering of background gas particles~\cite{schlosshauer-selbach_decoherence_2008,PhysRevA.84.052121} can be neglected whenever $ \gammagas \totalTime \ll 1 $, where $\gammagas =16 \pi \sqrt{2\pi} \pressure \radius^2/ [\sqrt{3} \massgas \vgas]$  is the gas scattering rate with $\radius$ the nanoparticle's radius, $\pressure$ the gas pressure, and $\massgas$ and $\vgas$ the averaged mass and thermal velocity of a gas particle, respectively. It is one of the key aspects of our proposal that the loop protocol is fast enough to easily guarantee $ \gammagas \totalTime \ll 1 $ at ultra-high vacuum $\pressure\sim10^{-9}~\text{mbar}$~\cite{Dimity_Tracy_loading}, that is, that the probability for a single scattering event in each experimental run is negligible.
The dynamics induced by \eqnref{eq:ME} lead to closed equations of motion for the first and second moments~\cite{SupplementalMaterial} that can thus be easily solved. Note, however, that the dissipator modelling frequency noise generates mixed non-Gaussian states. Hereafter we use expressions that are valid only for Gaussian states. In the presence of frequency noise, we have numerically checked that they are an excellent approximation in the relevant parameter regimes.
Non-harmonic terms in the potential might become relevant for large expansions. Experiments with optically levitated nanoparticles~\cite{gieseler_thermal_2013} show that these nonlinear contributions are not significant for expansions of up to $\eta\sim10^4$, where the spatial extent of the nanoparticle's center-of-mass position is of around $10$~\text{nm}. 
Furthermore, one could consider shallower traps to prevent nonlinearities provided the protocol is short enough such that $\gammagas \totalTime \ll 1$ and hence decoherence due to gas scattering is kept irrelevant.

\begin{figure}
    \centering
    \includegraphics[width=\columnwidth]{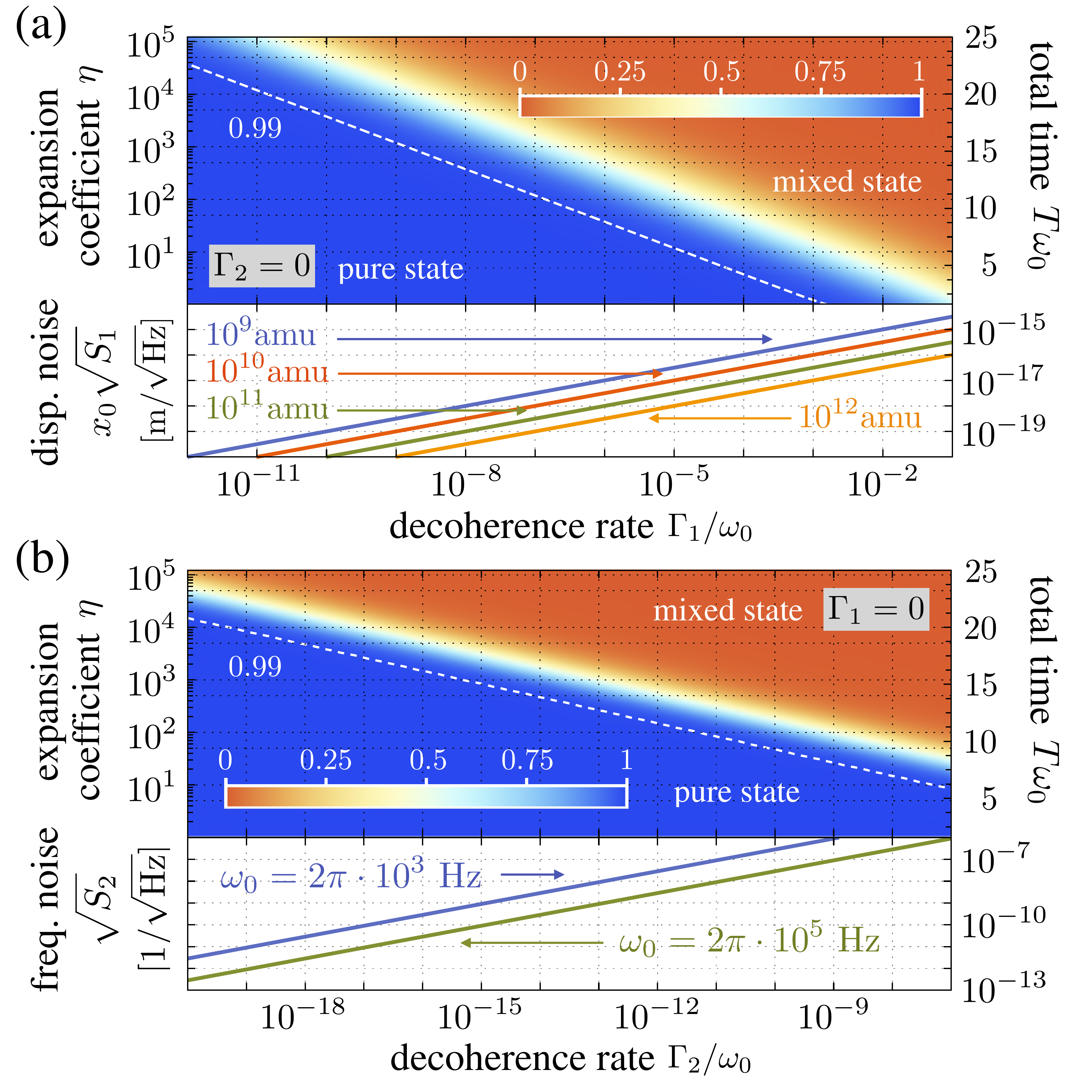}
    \caption{Coherent expansion. The purity at the end of the protocol as a function of the total protocol time $\totalTime$ and the decoherence rate (a) $\decoherence$ ((b) $\decoherenceFreq$). The lower panel in (a) relates the decoherence rate $\decoherence$ to the displacement noise spectrum $\displacementNoise$ for $\freq = 2\pi \times 10^5$Hz and several typical particle masses $\mass$.
    The lower panel in (b) relates the decoherence rate $\decoherenceFreq$ to the frequency noise spectrum $\frequencyNoise$ for two different frequencies (there is no mass-dependence for frequency noise). As the initial state we consider the ground state, $\thermalN=0$, with $\purity=1$. We considered a silica nanoparticle of $\radius =100~\text{nm}$ and mass density $\massdensity=2201~\text{kg}/\text{m}^{-3}$.}
    \label{fig:2}
\end{figure}

In \figref{fig:2}, we show the purity $\purity = \tr \spare{\densityMatrix^2(\totalTime)}$ of the state after the loop protocol, where $\densityMatrix(\totalTime)$ is calculated using \eqnref{eq:ME} with $\densityMatrix(0) = \ketbra{0}{0}$. In \figref{fig:2}(a) (\figref{fig:2}(b)) we consider the impact of displacement (frequency) noise only. A coherent expansion of $\eta=10^2$, namely a quantum delocalization over a length scale $100 \times \xzpf$ requires $\decoh_1/\freq \approx 10^{-7}$ and $\decoh_2/\freq \approx 10^{-11}$, that is a displacement noise $\xzpf \sqrt{\displacementNoise} \sim 10^{-18}\text{m}/\sqrt{\text{Hz}}$ and a frequency noise of $\sqrt{\frequencyNoise} \sim 10^{-8}/\sqrt{\text{Hz}}$. Figure \ref{fig:2} quantifies the experimental challenge (dechorence rate levels) required to explore macroscopic quantum physics. It might be convenient in order to reduce decoherence to consider protocols where the harmonic frequency in $\Hinv$ is smaller than in $\Hplus$. While this will make the protocol longer in time (larger $\totalTime$), it might be convenient if decoherence due to gas scattering is still negligible $ \gammagas \totalTime \ll 1 $. These modifications can be analyzed and optimized for a given experimental scenario in a systematic way using the results presented here. 
Figure \ref{fig:2} illustrates the fragility of macroscopic quantum physics. It quantifies how decoherence modelled by displacement and frequency noise limits the scale of quantum delocalization that can be achieved with a nanoparticle.
In this regard, since collapse models effectively induce displacement noise~\cite{PhysRevA.84.052121,RevModPhys.85.471}, the loop protocol can also be directly used to falsify them by measuring a high state purity.

Let us now show that the loop protocol also enhances static force sensing. Similarly to the experiment done with free falling nanoparticles in~\cite{Hebestreit_staticForceSensing_2018}, we consider the loop protocol in the presence of a static force $F$. That is, the Hamiltonians $\Hplusminus$ in the protocol are modified to
\begin{equation}
      \HplusminusF = \Hplusminus + \force \position= \Hplusminus + \hbar\freq \unitlessForce\unitlessPosition
    \label{eq:HamiltonianForce}.
\end{equation}
Here $\unitlessForce=\force\xzpf/(\hbar\freq)$ is the dimensionless static force to be detected. The state after the loop protocol will now depend on $\unitlessForce$, that is $\densityMatrix(\totalTime, \unitlessForce)$. Using standard methods in quantum metrology, we consider the quantum Fisher information~\cite{braunstein_statistical_1994} of  $\densityMatrix(\totalTime, \unitlessForce)$, denoted as $\QFI$, and the  Cram\'er-Rao bound, to calculate the minimal force $\unitlessForcemin \equiv 1/\sqrt{\QFI}$ that can be detected with the loop protocol. For a Gaussian state, $\QFI$ can be calculated from first and second order moments \cite{jiang_quantum_2014}, see \cite{SupplementalMaterial}.
\begin{figure}
    \centering
    \includegraphics[width=\columnwidth]{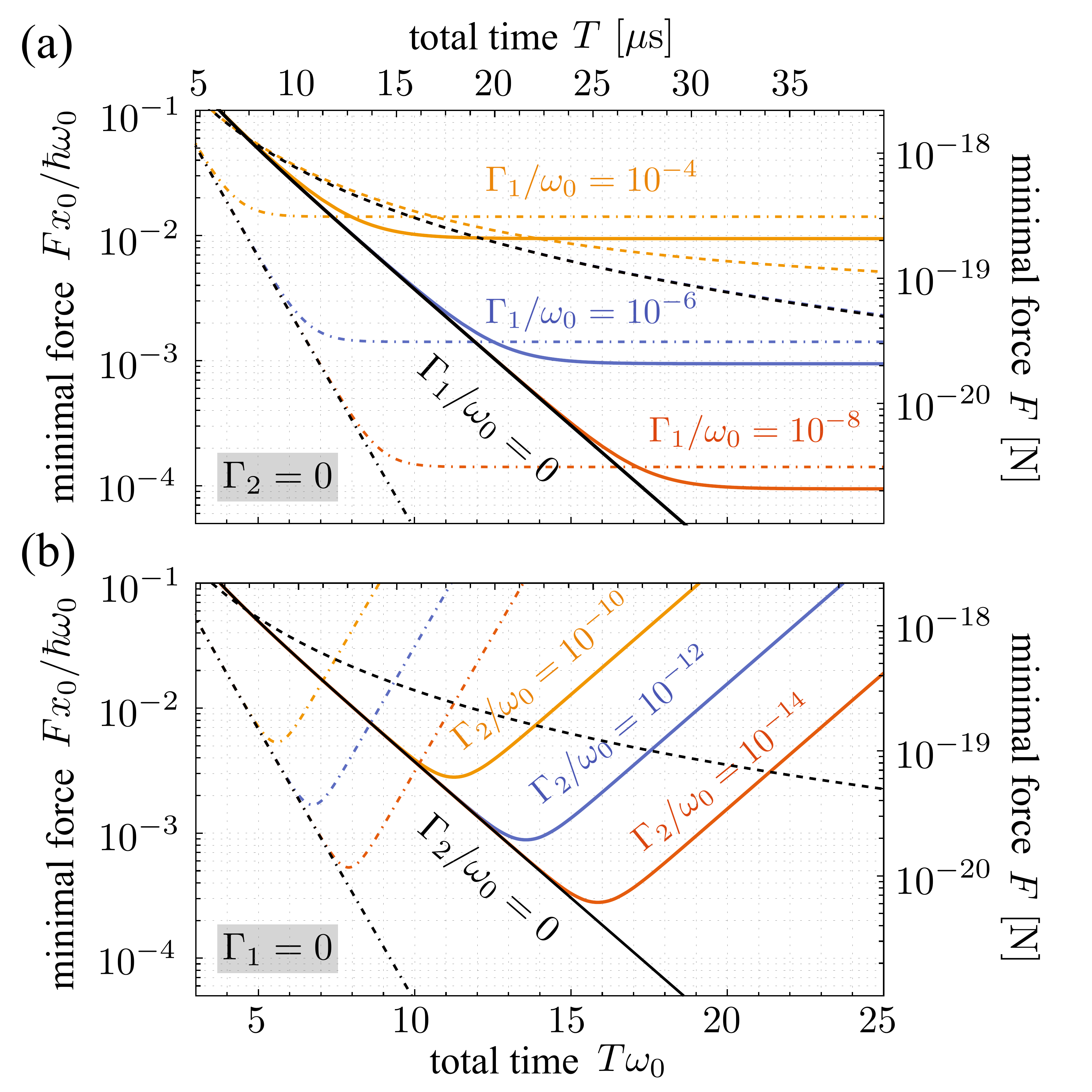}
    \caption{Force sensing. Minimal detectable force as a function of the total time $\totalTime$. Results are shown in the complete absence of decoherence (black lines) and in color for various values of (a) $\decoherence$ with $\decoherenceFreq=0$ ((b) $\decoherenceFreq$ with $\decoherence=0$). Solid lines show the performance of the loop protocol. Dashed (dash-dotted) lines show the performance of a particle that is evolving freely (in an inverted potential) for a time $\totalTime$. The right (top) axis shows the values in Newton (microseconds) for a silica nanoparticle of of $\radius =100~\text{nm}$ and $\freq = 2\pi \times 10^5$Hz.}
    \label{fig:3}
\end{figure}
In \figref{fig:3} we display $\unitlessForcemin$ as a function of $\totalTime$. The solid, dashed, and dotted-dashed black lines (the same in both panels) are $\unitlessForcemin$ in the absence of decoherence ($\decoh_1=\decoh_2=0$) for the loop protocol, free evolution ($\freq=0$), and evolution in a constant inverted potential, respectively. These lines show the enhanced force sensing, as compared to free dynamics~\cite{Hebestreit_staticForceSensing_2018}, provided by the fast and large expansion induced by an inverted potential.  In~\cite{SupplementalMaterial}, we provide the analytical expressions of $\unitlessForcemin$ in these three cases as well as the modified key equality \eqnref{eq:bangbang} of the loop protocol in the presence of a static force.  The colour lines in \figref{fig:3}(a) (\figref{fig:3}(b))  show $\unitlessForcemin$ in presence of displacement noise (frequency noise). Interestingly, displacement noise saturates $\unitlessForcemin$ as a function of time. In contrast, frequency noise provides an optimal time where $\unitlessForcemin$ is minimal. In both \figref{fig:2} and \figref{fig:3} one can see that, as expected from the form of the dissipators in \eqnref{eq:ME}, frequency noise becomes dominant whenever $\decoh_2 \eta^4 \gtrsim \decoh_1 \eta^2 $, that is, as soon as the nanoparticle expands to  regimes $ \eta \gtrsim \sqrt{\decoh_1 /\decoh_2} $.

Let us also show how the loop protocol can enhance the entanglement generation via a weak static interaction of two nanoparticles. We consider two particles, one with position vector $(\position_1,0,0)$ and the other with $(\position_2,d,0)$,  interacting via the general central potential 
\be \label{eq:Vint}
 \Vint = \frac{\scaleG}{\spare{(\position_1-\position_2)^2+\distanceD^2}^{\scaleR/2}}. 
 \ee 
We therefore assume that the motion of the two particles in their corresponding trapping potential, as described  by the position $\position_{1 (2)}$ and momentum $\momentum_{1 (2)}$ for particle $1$ ($2$), is parallel and separated by a constant distance $\distanceD$.  The dimensional real parameter $\scaleG$ and the dimensionless integer $\scaleR$ determine the type of interaction (\eg~Coulomb~\cite{Frimmer_ChargeNanoparticle_2017}, Casimir~\cite{Emig_CasimirForces_2007}, gravitational~\cite{Marletto_GravitationalEntanglement_2017, Bose_GravitationalEntanglement_2017}). By assuming $d^2 \gg \avg{(\position_1-\position_2)^2}$, one can Taylor expand \eqnref{eq:Vint} to get, together with \eqnref{eq:Hamiltonian}, the total quadratic Hamiltonian
\begin{equation}
  \frac{\Htwo}{\hbar\freqRe}= \frac{1}{4}\sum_{j=1,2} \pare{{\unitlessMomentum_j^2 \pm \unitlessPosition_j^2}}+\frac{\coupling}{\freqRe}\unitlessPosition_1\unitlessPosition_2.
    \label{eq:twoParticleHamiltonian}
\end{equation}
Here $\freqRe^2 \equiv \freq^2 + \scaleG \scaleR/(\distanceD^{\scaleR+2} \mass)$ is the shifted harmonic oscillator and $g \equiv  \scaleG \scaleR/(\distanceD^{\scaleR+2} 2 \mass \freqRe)$ the coupling rate. The dimensionless position and momentum operators are defined with the zero-point motion that depends on the shifted frequency $\freqRe$. From \eqnref{eq:twoParticleHamiltonian} one can realize that the coupling rate between nanoparticles in a center-of-mass state which is expanded by a factor $\eta$, as it is achieved in the loop protocol, will be enhanced to $\eta^2 \coupling$. In the loop protocol, the state is in the expanded phase during an amount of time $\timeHarmonic$. To generate entanglement one requires  $\eta^2 \coupling \timeHarmonic>1 $, a condition that profits from the the $\eta^2$ enhancement.
\begin{figure}[t]
    \centering
    \includegraphics[width=\columnwidth]{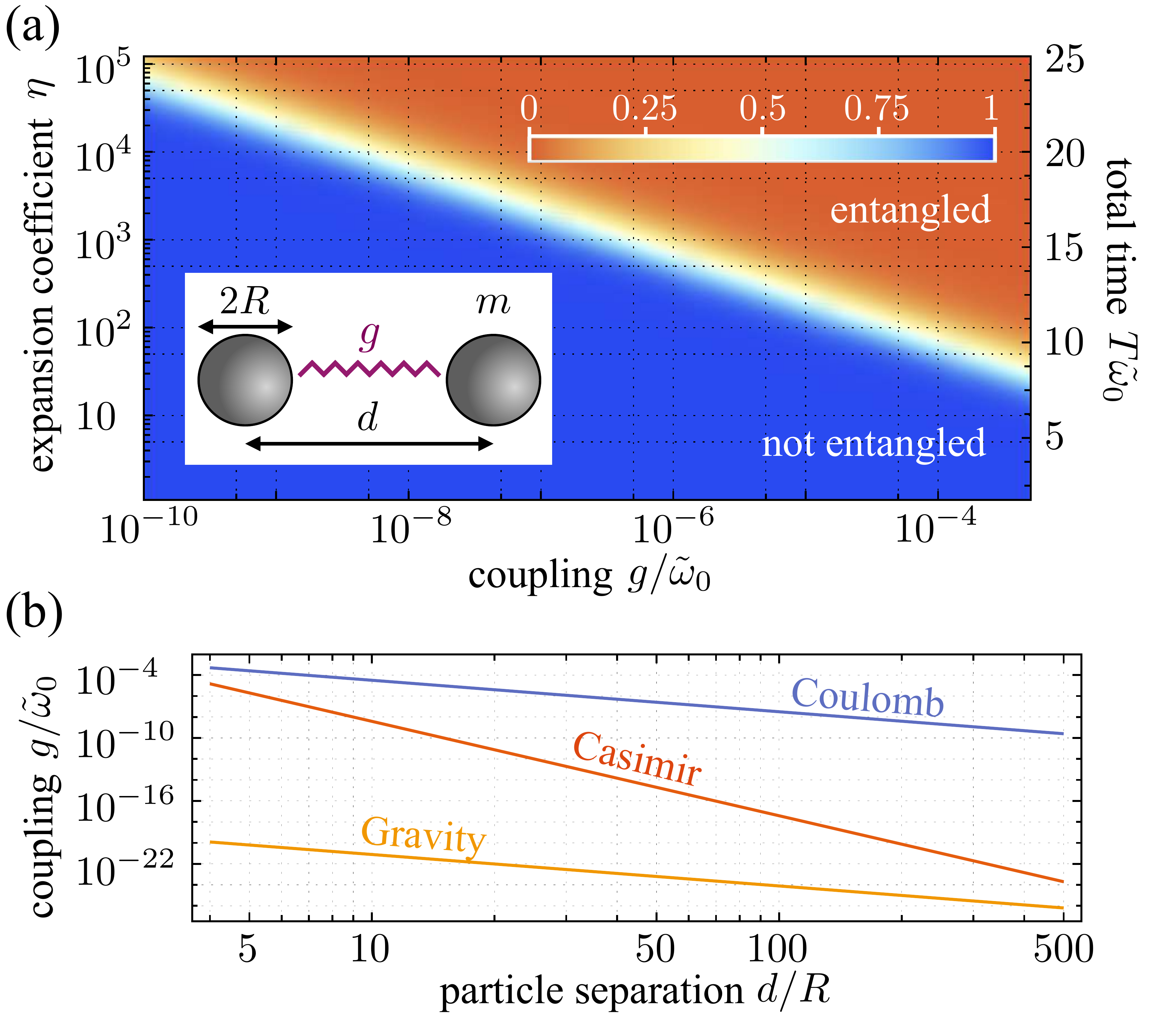}
    \caption{Entanglement via weak forces: (a) The purity at the end of the protocol as a function of the protocol time $\totalTime$ and coupling strength $\coupling$. In the absence of decoherence, the decrease of purity signals the creation of entanglement (starting from the ground state, $\thermalN=0$, with initial $\purity=1$). (b) Coupling $\coupling$ as a function of the particle separation $\distanceD/\radius$ for Coulomb, Casimir and gravitational interaction, for a silica nanoparticle of $\radius =100~\text{nm}$ and $\freqRe = 2\pi \times 10^5$Hz. In the case of Coulomb interaction, each particle is assumed to carry a single electron charge. See~\cite{SupplementalMaterial} for further details.}
    \label{fig:4}
\end{figure}
To confirm this reasoning, let us assume that the initial state is the product state $\densityMatrix (0) = \ket{0}_1\bra{0} \otimes  \ket{0}_2\bra{0}$ and that the evolution during the loop protocol is coherent. In Fig.~\ref{fig:4}(a) we show the purity $\purity = \tr \spare{\densityMatrix_1^2(\totalTime)}$ of the reduced state of  particle 1, namely $\densityMatrix_1 (\totalTime) = \tr_2 \spare{\densityMatrix(\totalTime)}$ (here $ \tr_2 [\cdot]$ is the partial trace over particle 2). Since the initial state is a product of pure states and the evolution is coherent, the reduced purity at the end of the loop protocol is a measure of the entanglement generated between the two particles. As an example, if the loop can be done coherently up to $\eta = 10^3$ (see \figref{fig:2} for the required decoherence rates), entanglement could be generated with a weak coupling rate of $\coupling \approx 2\pi \times 0.1~\text{Hz}$ in the $\totalTime \approx 25~\mu\text{s}$ that the loop protocol requires. As shown in Fig.~\ref{fig:4}(b), this coupling rate corresponds to the Coulomb interaction of two nanoparticles of $\radius=100~\text{nm}$, each having a single charge, separated by a distance $\distanceD\approx 3.2~\mu\text{m}$. Entangling two nanoparticle interacting with such a week coupling rate would be a proof-of-principle demonstration of the enhancement induced by large expansions. Entangling via Casimir interaction~\cite{Emig_CasimirForces_2007} would offer a novel approach to measure these weak forces. Entangling via gravity would have fundamental implications regarding the quantum character of gravity~\cite{belenchia_quantum_2018}. However, as shown in Fig.~\ref{fig:4}b,  the gravitational interaction leads to a coupling rate of the order of $\coupling/\freqRe \sim 10^{-23}$, which shows that entangling via gravitational force is a formidable task.

In summary, we have proposed a loop protocol to delocalize the center-of-mass state of a nanoparticle over large distances and compress it back to the initial localized state. The state in the expanded phase is very fragile to its environment. The loop protocol can thus be used to detect the environment, either its noisy signals, the presence of a static weak force, or the coherent interaction with another nanoparticle. We have presented the results in a way that one can easily extract the required levels of decoherence and noise in order to explore macroscopic quantum physics. From the current state of the art in experiments, where ground-state cooling of nanoparticles has been recently achieved, our results provide a quantitative path to progressively venture into the exciting regime of macroscopic quantum physics.

TW and MRL contributed equally to this work. We thank P. Feldmann, D. Giannandrea, C. Gonzalez-Ballestero, N. Kiesel, and A. Serafini for helpful discussions.
ORI and MA thank the hospitality of the Perimeter Institute in 2017, where they had the first discussions about this project.
This project has received funding from the European Research Council (ERC) under the European Union’s Horizon 2020 research and innovation programme (grant agreement No. [951234]).
TW acknowledges financial support from the Alexander von Humboldt foundation.
ET acknowledges financial support from Project PGC2018-094792-B-I00 (MCIU/AEI/FEDER,UE), CSIC Research Platform PTI-001, and CAM/FEDER Project No. S2018/TCS-4342 (QUITEMAD-CM). 

Note Added. During the submission of this manuscript, the preprint arXiv:2012.07815 \cite{cosco_enhanced_2020} discussing related ideas has been posted.

\bibliography{bibliography.bib}

\newpage
\onecolumngrid
\setcounter{figure}{0}
\setcounter{equation}{0}
\renewcommand{\thefigure}{S\arabic{figure}}
\renewcommand{\theequation}{S\arabic{equation}}
\renewcommand{\bibnumfmt}[1]{[S#1]}
\renewcommand{\citenumfont}[1]{S#1}
\setcounter{section}{0}
\renewcommand\thesection{S\arabic{section}}

\begin{center}
    \vspace{1cm}
	\large{\bf{Supplemental Material}}
\end{center}

In this Supplemental Material we provide more details. 
In Sec.~\ref{SMsec:unitaries}, we derive the key equality (Eq. (2) in the Letter) in the presence of a static force.
In Sec.~\ref{SMsec:MasterEquation} we summarize the derivation based on a thermal bath of the master equation used in this work.
In Sec.~\ref{SMsec:eoms} we present the explicit expressions for the equations of motion.
In Sec.~\ref{SMsec:SwitchingTimeError} we show that having errors in the duration of the harmonic phase of the protocol is equivalent to frequency noise and that therefore it is already included in our analysis via $\decoherenceFreq$.
In Sec.~\ref{SMsec:QFI} we recall how to calculate the quantum Fisher information $\QFI$ from the first and second order moments and give explicit expressions for the scenarios considered in the main text in the absence of decoherence.
In Sec.~\ref{SMsec:entanglement} we provide more details on how to derive the interaction strength $\coupling$ between two levitated particles for the gravitational, Coulomb, and Casimir interaction. We also state the explicit expressions for these three forces.

\section{Derivation of the loop unitary descriptions}\label{SMsec:unitaries}
In the absence of decoherence and in the presence of a static force, the loop protocol is described by the product of the following three time evolution unitary operators
\begin{equation}
    \UloopF = \exp\pare{\frac{-i \HminusF \timeInverted}{\hbar}} \exp\pare{\frac{-i \HplusF \timeHarmonic}{\hbar}} \exp\pare{\frac{-i \HminusF \timeInverted}{\hbar}}
    \equiv \UinvertedF(\timeInverted) \UharmonicF(\timeHarmonic) \UinvertedF(\timeInverted).
    \label{SMeq:UloopF}
\end{equation}
We show here that for $\timeHarmonic = (2l+1)\pi/(2\freq)$, with $l$ an integer, the whole loop protocol has an effective unitary description given by
\begin{equation}
    \UloopF = \hat{D}(\shiftF) \Uharmonic(\timeHarmonic) e^{i \phase(\unitlessForce,\timeInverted)}.
    \label{SMeq:UloopF_toProve}
\end{equation}
Here, $\Dop(\shiftF) \equiv\exp\pare{ i \spare{ \Im(\shiftF)\unitlessPosition - \Re(\shiftF) \unitlessMomentum}} $ denotes the displacement operator, $\shiftF= \unitlessForce \pare{ 1 - 2 e^{\timeInverted\freq} }(1+i)$, and $\phase(\unitlessForce,\timeInverted)$ is a global phase.
Note that the case $f=0$ in \eqnref{SMeq:UloopF_toProve} corresponds to Eq.~(2) in the main text.

Defining a shifted position operator $\shiftedX = \unitlessPosition + 2\unitlessForce$ and ignoring constant terms simplifies $\HplusminusF$ to 
\begin{equation}
    \HplusF = \frac{\hbar\freq}{4}\pare{\shiftedX^2+\unitlessMomentum^2} \qquad\text{ and }\qquad 
    \HminusF= \frac{\hbar\freq}{4}\pare{ (-\shiftedX^2+\unitlessMomentum^2) + 8 \unitlessForce \shiftedX}.
\end{equation}
The proof then follows by making repeated use of the standard Baker-Campbell-Hausdorff (BCH) formula
\begin{equation}
    e^{\Aop} \Bop e^{-\Aop} = \BathOperator + \spares{\Aop,\Bop} + \frac{1}{2!} \spares{\Aop,\spares{\Aop,\Bop}} + \cdots = \sum_{n=0}^\infty \frac{\mathcal{L}(\Aop)^n \Bop}{n!}
\end{equation}
where $\mathcal{L}(\Aop)\Bop\equiv \spares{\Aop,\Bop}$, and the fact that the shifted position operator still fulfills the canonical commutation relation $\com{\shiftedX}{\unitlessMomentum} = 2i$.
We divide the derivation in three steps, starting from Eq.~(\ref{SMeq:UloopF}) with an added identity on the right:
\begin{equation}
    \UloopF = 
    \underbrace{\overbrace{\UinvertedF(\timeInverted)
    \underbrace{\UharmonicF(\timeHarmonic) \UinvertedF(\timeInverted)
    \UharmonicF{}^\dagger(\timeHarmonic)}_\text{step 1}}^\text{step 2}
    \UharmonicF(\timeHarmonic)}_\text{step 3}.
\end{equation}
In the following, we assume $\timeHarmonic = (2l+1)\pi/(2\freq)$, with $l$ an integer and neglect all global phases.
\\In step 1, one finds analytical expressions for $\mathcal{L}(\HplusF)^n \HminusF$ which can be summed explicitly to obtain
\begin{equation}
    \UharmonicF(\timeHarmonic)\UinvertedF(\timeInverted)\UharmonicF{}^\dagger(\timeHarmonic) =
    \exp\pare{i \frac{\timeInverted\freq}{4} \spare{ \pare{-\shiftedX^2+\unitlessMomentum^2} + 8 \unitlessForce \unitlessMomentum} }.
\end{equation}
In step 2, one multiplies the above result with $\UinvertedF$ from the left. In this case, it possible to use a special case of the BCH formula \cite{SM-van-brunt_special-case_2015} which yields
\begin{equation}
    \UinvertedF(\timeInverted)
    \UharmonicF(\timeHarmonic) \UinvertedF(\timeInverted)
    \UharmonicF{}^\dagger(\timeHarmonic)= 
    \exp \spare{2 i \unitlessForce \pare{ \shiftedX-\unitlessMomentum} \pare{ 1 - e^{\timeInverted\freq} }} = \Dop\pare{2\unitlessForce \spare{ 1 - e^{\timeInverted\freq} }[1+i]} \equiv \Dop(\helpshiftF)
\end{equation}
In step 3, one makes use of the fact that
$\UharmonicF(\timeHarmonic) = \Dop(-\unitlessForce) \Uharmonic(\timeHarmonic) \Dop(\unitlessForce)$ to write
\begin{equation}
    \UloopF = \Dop(\helpshiftF)
    \Dop(-\unitlessForce)\Uharmonic(\timeHarmonic)\Dop(\unitlessForce).
\end{equation}
Finally, using the BCH formula to swap the last two terms one obtains the final result
\begin{equation}
    \UloopF = \Dop(\helpshiftF)
    \Dop(-\unitlessForce)\Dop(-i \unitlessForce)\Uharmonic(\timeHarmonic) = \Dop\pare{\spare{1-2e^{\timeInverted\freq}}[1+i]}
    \Uharmonic(\timeHarmonic).
\end{equation}

\section{Master equation derivation} \label{SMsec:MasterEquation}

We consider our system coupled linearly to a bath as specified by the Hamiltonian 
\begin{equation}
    \Hop = \systemHamiltonian + \bathHamiltonian - \genericOperator \BathOperator,
\end{equation}
where $\bathHamiltonian$ is the Hamiltonian of the free bath and $\genericOperator$ ($\BathOperator$) is a generic system (bath) operator.
Using the Born-Markov approximation, one can show that the master equation for the system is given by \cite{SM-breuer_theory_2010}
\begin{equation}
    \frac{\partial\rhoop(t)}{\partial t} = - \frac{i}{\hbar} \com{\systemHamiltonian}{\rhoop(t)}
    + \frac{1}{\hbar^2} \int_0^\infty d\tau \pare{ 
        \frac{i}{2} \dissipationKernel(\tau) \com{\genericOperator}{\cpare{\genericOperator(-\tau),\rhoop(t)}}
        - \frac{1}{2} \noiseKernel(\tau) \com{\genericOperator}{\spare{\genericOperator(-\tau),\rhoop(t)}}
    },
    \label{SMeq:born_markov_master_equation}
\end{equation}
where $\genericOperator(-\tau) = \exp\pare{-i \freeHamiltonian \tau/\hbar} \genericOperator \exp\pare{i \freeHamiltonian\tau/\hbar}$ with $\freeHamiltonian = \systemHamiltonian + \bathHamiltonian$. Moreover, we defined the dissipation and noise kernels
\begin{align}
    \dissipationKernel(\tau) &\equiv i \langle \com{B}{B(-\tau)} \rangle = 2\hbar \int_0^\infty d\omega \noiseJ \sin\pare{\omega\tau} \\
    \noiseKernel(\tau) &\equiv \langle \cpare{B,B(-\tau)} \rangle = 2\hbar \int_0^\infty d\omega \noiseJ \coth(\hbar\omega/2k_B \temperature)\cos\pare{\omega\tau}
\end{align}
where the brackets denote trace over the bath.
All the information concerning the bath is thus incorporated in the function $\noiseJ$.
In the case of an harmonic oscillator $\systemHamiltonian = \Hplus$ with frequency $\freq$ and $\genericOperator = \unitlessPosition$ or $\genericOperator = \unitlessPosition^2$ one has
\begin{equation}
    \unitlessPosition(-\tau) = \cos \pare{\freq\tau}\, \unitlessPosition - \sin \pare{\freq\tau}\, \unitlessMomentum \quad\text{or}\quad
    \unitlessPosition^2(-\tau) = \frac{1 + \cos \pare{2\freq\tau}}{2} \unitlessPosition^2 + \frac{1 - \cos \pare{2\freq\tau}}{2} \unitlessMomentum^2 - \sin \pare{2\freq\tau} \xppx.
\end{equation}
Therefore, the last term in the master equation in Eq.~\eqref{SMeq:born_markov_master_equation} will contain several terms, with coefficients obtained performing the corresponding integrals.
At this point we consider Ohmic baths with a Lorentzian cutoff, namely
\begin{equation}
    J_{\nu}(\omega) = \frac{2 \mass \gamma_\nu}{\pi}\omega \frac{\cutOff^2}{\cutOff^2+\omega^2}
\end{equation}
where $\noiseRate_\nu$ refers to $\genericOperator = \unitlessPosition$ ($\nu=1$) and $\genericOperator = \unitlessPosition^2$ ($\nu=2$) respectively.
Working in the high temperature limit, $k_B \temperature\rightarrow\infty$, $\noiseRate_{\genericOperator}\rightarrow 0$ but with $k_B \temperature \noiseRate_\nu$ kept constant,
allows one to compute the aforementioned integrals.
Most of the terms will vanish in this limit, yielding the following master equation
\begin{equation}
    \frac{\partial\rhoop(t)}{\partial t} = 
    - \frac{i}{\hbar} \com{\systemHamiltonian}{\rhoop(t)}
    - \frac{2m \noiseRate_1 k_B \temperature}{\hbar^2} \com{\unitlessPosition}{\spare{\unitlessPosition,\rhoop(t)}}
    - \frac{2m \noiseRate_2 k_B \temperature}{\hbar^2} \com{\unitlessPosition^2}{\spare{\unitlessPosition^2,\rhoop(t)}}.
    \label{SMeq:calderia-legget-master-equation}
\end{equation}
which is analogous to the the high temperature limit of the Caldeira-Legget master equation \cite{SM-breuer_theory_2010} but also including $\unitlessPosition^2$ noise.
For an inverted harmonic oscillator $\systemHamiltonian = \Hminus$ one has
\begin{equation}
    \unitlessPosition(-\tau) = \cosh \pare{\freq\tau}\, \unitlessPosition - \sinh \pare{\freq\tau}\, \unitlessMomentum \quad\text{or}\quad
    \unitlessPosition^2(-\tau) = \frac{1 + \cosh \pare{2\freq\tau}}{2} \unitlessPosition^2 + \frac{1 - \cosh \pare{2\freq\tau}}{2} \unitlessMomentum^2 - \sinh \pare{2\freq\tau} \xppx.
\end{equation}
This changes the integrals involved in the derivation and one has to take their principal values.
Nevertheless, in the high temperature limit, one ends up with the same result as above Eq.~\eqref{SMeq:calderia-legget-master-equation}.

\section{Equations of motion}\label{SMsec:eoms}

The equations of motion for the first and second order moments corresponding to Hamiltonian (\ref{eq:HamiltonianForce}) in the main text also including displacement and frequency noise are

\begin{equation}\label{eq:eomsWithForce}
        \begin{aligned}
        \expect{\dot{\unitlessPosition}^2} &= 2 \expect{\xppx}\\
        \expect{\dot{\unitlessMomentum}^2} &=\mp 2 \expect{\xppx}+8\frac{\decoherence}{\freq}+32\frac{\decoherenceFreq}{\freq}\expect{\unitlessPosition^2} -4\unitlessForce\expect\unitlessMomentum\\
        \expect{\dot{\xppx}} &=\expect{\unitlessMomentum^2}\mp\expect{\unitlessPosition^2} -2\unitlessForce\expect\unitlessPosition \\
        \expect{\dot{\unitlessPosition}} &= \expect{\unitlessMomentum}\\
        \expect{\dot{\unitlessMomentum}} &= \mp\expect{\unitlessPosition} - 2\unitlessForce.
    \end{aligned}
\end{equation}
Here, the upper (lower) sign corresponds to the motion in a harmonic (inverted) potential and $\expect\xppx = (\expect{\unitlessPosition\unitlessMomentum}+\expect{\unitlessMomentum\unitlessPosition})/2$. 

The equations of motion of two coupled levitated particles, corresponding to Hamiltonian (\ref{eq:twoParticleHamiltonian}) of the main text, are
\begin{equation}
    \begin{aligned}
        \expect{\dot{\unitlessPosition}_j^2} &= 2 \expect{\xppx_j}\\
        \expect{\dot{\unitlessMomentum}_j^2} &=\mp 2 \expect{\xppx_j}-4\frac{\coupling}{\freqRe}\expect{\unitlessMomentum_j\unitlessPosition_k}\\
        \expect{\dot{\xppx}_j} &=\expect{\unitlessMomentum_j^2}\mp\expect{\unitlessPosition_j^2}-2\frac{\coupling}{\freqRe}\expect{\unitlessPosition_1\unitlessPosition_2}\\
        \dot{\expect{\unitlessMomentum_j\unitlessPosition_k}} &= \mp\expect{\unitlessPosition_j\unitlessPosition_k}+\expect{\unitlessMomentum_j\unitlessMomentum_k}-2\frac{\coupling}{\freqRe}\expect{\unitlessPosition_k^2}\\
        \dot{\expect{\unitlessPosition_1\unitlessPosition_2}} &= \expect{\unitlessMomentum_1\unitlessPosition_2}+\expect{\unitlessPosition_1\unitlessMomentum_2}\\
        \dot{\expect{\unitlessMomentum_1\unitlessMomentum_2}} &= \mp\expect{\unitlessPosition_1\unitlessMomentum_2}\mp\expect{\unitlessMomentum_1\unitlessPosition_2}-2\frac{\coupling}{\freqRe}\left(\expect{\xppx_1}+\expect{\xppx_2}\right),
    \end{aligned}
\end{equation}
where $j=1,2$ and $k=1,2$ such that $j\neq k$. The first order moments decouple from these second-order equations of motion even in the presence of the interaction $\coupling$ and their solutions remain zero if their initial values are zero, i.e. $\expect{\unitlessPosition}=\expect{\unitlessMomentum}=0$, and therefore they are omitted here.

\section{Errors in switching times} \label{SMsec:SwitchingTimeError}

Let us discuss the impact of having some noise in $\timeHarmonic$, which is the part of the protocol where the state is most fragile. We will show that these errors can be included into our frequency noise quantified by $\decoherenceFreq$.

We assume that in each experimental run the time spent in the harmonic potential is given by $\timeHarmonic + \timeError$. In the absence of any other noise source, for a particular value of $\timeError$ the state at the end of the protocol is pure and Gaussian. We denote this state by $\ket{\psi_\timeError(\totalTime)}$. Assuming $\ket{\psi_\timeError(0)} = \ket{0}$, the second order moments of $\ket{\psi_\timeError(\totalTime)}$ are given by
\begin{align}
    \expect{\unitlessPosition^2}_\timeError &\equiv \bra{\psi_\timeError(\totalTime)} \unitlessPosition^2 \ket{\psi_\timeError(T)} =
    \cosh ^2(2 \timeInverted\freq) - \sinh (2 \timeInverted\freq) (\sinh (2 \timeInverted\freq) \cos (2 \timeError\freq)+\sin (2 \timeError\freq)) \\
    \expect{\unitlessMomentum^2}_\timeError &\equiv \bra{\psi_\timeError(\totalTime)} \unitlessMomentum^2 \ket{\psi_\timeError(T)} =
    \cosh ^2(2 \timeInverted\freq) - \sinh (2 \timeInverted\freq) (\sinh (2 \timeInverted\freq) \cos (2 \timeError\freq)-\sin (2 \timeError\freq)) \\
    \expect{\xppx}_\timeError &\equiv \bra{\psi_\timeError(\totalTime)} \xppx \ket{\psi_\timeError(T)} = \sinh (4\timeInverted\freq) \sin^2(\timeError\freq)
\end{align}
We assume $\timeError$ to be a stochastic error sampled from a Gaussian distribution with variance $\timeErrorVariance$. The ensemble averaged mixed state is thus given by
\begin{equation}
    \rhoop_\timeErrorVariance(T) \equiv 
    \int_{-\infty}^{\infty} d\timeError\; P_\timeErrorVariance(\timeError) \ket{\psi_\timeError(T)}\bra{\psi_\timeError(\totalTime)} =
    \int_{-\infty}^{\infty} d\timeError\; \frac{1}{\sqrt{2\pi}\timeErrorVariance} \exp\pare{-\frac{\timeError^2}{2 \timeErrorVariance^2}} \ket{\psi_\timeError(T)}\bra{\psi_\timeError(\totalTime)}.
\end{equation}
Its second order moments can be computed via
\begin{equation}
    \expect{\unitlessPosition^2}_\timeErrorVariance = \Tr\spare{\rhoop_\timeErrorVariance \unitlessPosition^2} = 
    \int_{-\infty}^{\infty} d\timeError\; P_\timeErrorVariance(\timeError) \bra{\psi_\timeError(\totalTime)} \unitlessPosition^2 \ket{\psi_\timeError(T)} =
    \int_{-\infty}^{\infty} d\timeError\; \frac{1}{\sqrt{2\pi}\timeErrorVariance} \exp\pare{-\frac{\timeError^2}{2 \timeErrorVariance^2}} \expect{\unitlessPosition^2}_\timeError
\end{equation}
and analogously for $\expect{\unitlessMomentum^2}_\timeErrorVariance$ and $\expect{\xppx}_\timeErrorVariance$. One obtains
\begin{align}
    \expect{\unitlessPosition^2}_\timeErrorVariance = \expect{\unitlessMomentum^2}_\timeErrorVariance = 
    \frac{1}{2} \left(1 -2 e^{-2 \timeErrorVariance^2\freq^2} \sinh ^2(2 \timeInverted\freq)+\cosh (4 \timeInverted\freq)\right) \quad\text{and}\quad
    \expect{\xppx}_\timeErrorVariance = \frac{1}{2} \left(1-e^{-2 \timeErrorVariance^2\freq^2}\right) \sinh (4 \timeInverted\freq).
\end{align}

Alternatively, we can consider the second order moments of the state obtained by solving our equations of motion in presence of frequency noise only.
Doing so one finds that, for $\timeErrorVariance\freq \ll 1$ and $\timeInverted\freq \gg 1$, the moments obtained with frequency noise and time errors are the same, provided one makes the following identification
\begin{equation}
    \decoherenceFreq = \frac{\timeErrorVariance^2 \freq^3}{20+6 \pi}.
\end{equation}

\section{The quantum Fisher information for a single-particle Gaussian state}\label{SMsec:QFI}

For a single-particle Gaussian state the quantum Fisher information can be expressed as \cite{SM-jiang_quantum_2014}
\begin{equation}
    \QFI = \frac{\tr\spares{\symplecticEV^4\pare{\dot \covarianceMatrix \covarianceMatrix^{-1}}^2 - \pare{\dot \covarianceMatrix \auxMatrix}^2}}{2(\symplecticEV^4-1)}  
    + \dot{\boldsymbol{\meanVector}}^\dagger\covarianceMatrix^{-1} \dot{\boldsymbol{\meanVector}}.
    \label{eq:quantum_fisher_info}
\end{equation}
Here the derivatives need to be taken with respect to the dimensionless force $\unitlessForce$. The covariance matrix $\covarianceMatrix$ is given by
\begin{equation}
    \covarianceMatrix = \begin{pmatrix}
        \expect{\unitlessPosition^2} - \expect{\unitlessPosition}^2 &  \frac{\expect{\unitlessPosition\unitlessMomentum}+\expect{\unitlessMomentum\unitlessPosition}}{2} - \expect{\unitlessPosition}\expect{\unitlessMomentum} \\
        \frac{\expect{\unitlessPosition\unitlessMomentum}+\expect{\unitlessMomentum\unitlessPosition}}{2} - \expect{\unitlessPosition}\expect{\unitlessMomentum} & \expect{\unitlessMomentum^2} - \expect{\unitlessMomentum}^2
    \end{pmatrix}
    \equiv \begin{pmatrix} \varianceX & \covariance  \\ \covariance &\varianceP \end{pmatrix},
\end{equation}
$\boldsymbol\meanVector = \pare{ \expect{\unitlessPosition} , \expect{\unitlessMomentum}}^\dagger$ is the vector of means, $\auxMatrix =  \begin{pmatrix} 0 & 1  \\ -1 & 0 \end{pmatrix}$, and $\symplecticEV$ is the symplectic eigenvalue of $\covarianceMatrix$ given by $\symplecticEV=\sqrt{\varianceX \varianceP - \covariance^2}$.

In the absence of noise, the quantum Fisher information in the three scenarios considered in the main text (loop protocol, inverted potential, and free particle) is given by
\begin{equation}
    \begin{aligned}
    \QFI_\text{loop} = \frac{4 \left(1-2 e^{\frac{\totalTime\freq }{2}-\frac{\pi }{4}}\right)^2}{2 \thermalN+1},\qquad
    \QFI_\text{inv} = \frac{8 \sinh ^2\left(\frac{\totalTime\freq }{2}\right) \cosh (\totalTime\freq )}{2 \thermalN+1},\qquad
    \QFI_\text{free} = \frac{\totalTime^2\freq^2 \left(\totalTime^2\freq^2+4\right)}{4 \thermalN+2}.
    \end{aligned}
\end{equation}

\section{Entanglement via weak interaction: derivation of the coupling strength}\label{SMsec:entanglement}

\begin{figure}
    \centering
    \includegraphics[width=\columnwidth]{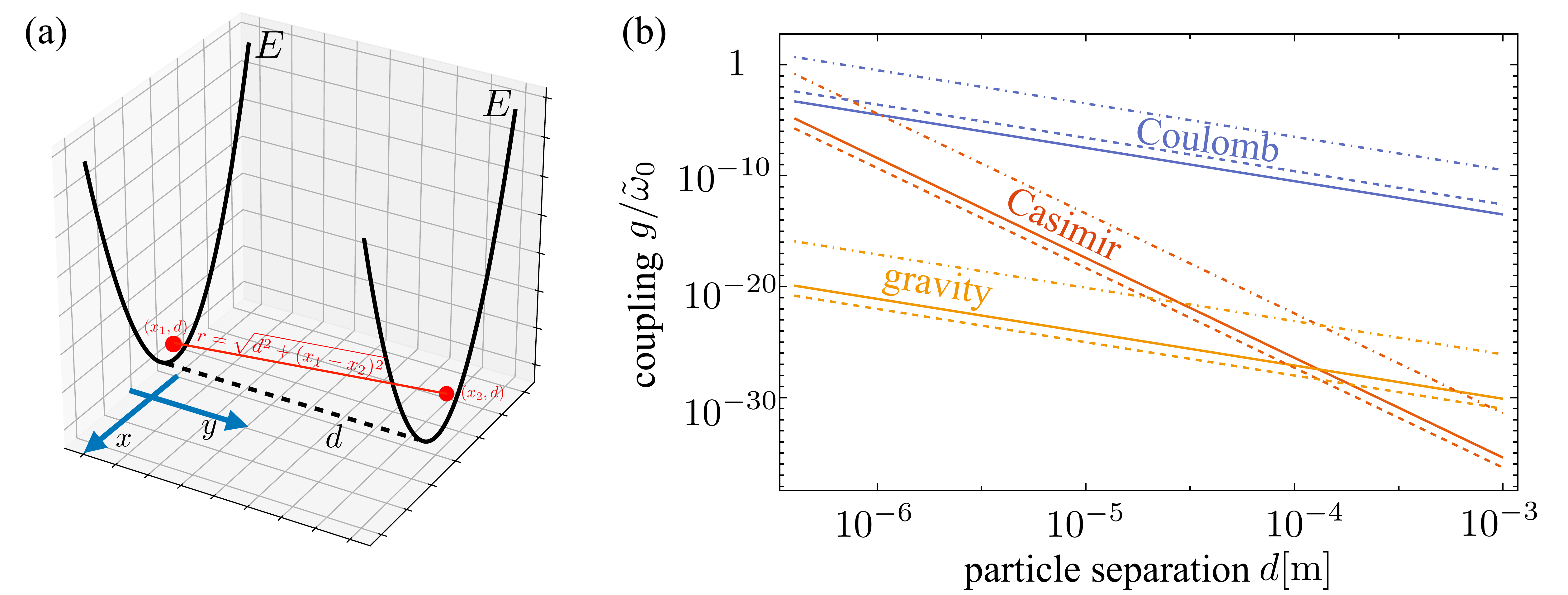}
    \caption{(a) Sketch of the two-particle scenario. Two levitated particles move in their respective trapping potential (here a harmonic one). The constant distance between the traps is $\distanceD$. The particles motion is confined to the $x$-direction, thus the distance between them is $r=\sqrt{\distanceD^2+(x_1-x_2)^2}$. (b) The coupling strength $\coupling/\freqRe$ due to the gravitational (yellow), Coulomb (blue), and Casimir (red) interaction as a function of the distance $\distanceD$ between the particles. For the Coulomb interaction it is assumed that each particle carries a single electron charge. Solid lines show the values for $\freqRe =2\pi\times10^5$ Hz and $\radius=100$ nm. Dashed lines for $\freqRe =2\pi\times10^5$ Hz and $\radius=50$ nm. Dot-dashed lines for $\freqRe =2\pi\times10^3$ Hz and $\radius=100$ nm.}
    \label{fig:SUP} 
\end{figure}

We consider two levitated particles, each moving in its respective trap (either a harmonic or inverted potential) parallel to each other. The particles interact via a weak force that has a $1/\distance^\scaleR$-dependence on a Hamiltonian level, with $\distance = \sqrt{(\position_1-\position_2)^2+\distanceD^2}$ the distance between the two particles center of mass position and $a$ a positive integer. Here, $\distanceD$ denotes the constant distance in the direction perpendicular to the motion in the trapping potential, and $\position_{1 (2)}$ is the position in the direction of motion of particle $1$ ($2$), see sketch in Fig.~\ref{fig:SUP} (a). A general expression for this type of entanglement-generating interaction can be obtained by Taylor-expanding

\begin{equation}
    \frac{1}{\distance^\scaleR}\approx \frac{1}{\distanceD^\scaleR}\left(1-\frac{\scaleR}{2}\frac{(\position_1-\position_2)^2}{\distanceD^2}\right).\\
\end{equation}
Neglecting the constant term and incorporating the terms proportional to $\position_j^2$ as a frequency shift, $\freq \rightarrow \freqRe$, leads to the total two-particle Hamiltonian (\ref{eq:twoParticleHamiltonian}) in the main text. 

More concretely, considering the gravitational, Coulomb, and Casimir\cite{SM-Emig_CasimirForces_2007} interaction between the two levitated particles at a distance $\distanceScalar$ we start from the respective potential energy,

\begin{equation}
    \begin{aligned}
        \potentialG(\distanceScalar) = -\frac{\gravConst\mass^2}{\distanceScalar}, \qquad
        \potentialC(\distanceScalar) = \frac{1}{4\pi\permittivity}\frac{\charge_1\charge_2}{\distanceScalar}, \qquad
        \potentialCa(\distanceScalar) = -\frac{\hbar \speedoflight}{\pi}\frac{23}{4}\left(\frac{\relativePermittivity-1}{\relativePermittivity+2}\right)^2\frac{\radius^6}{\distanceScalar^7}.
    \end{aligned}
\end{equation}
Here, $\gravConst$ is the gravitational constant, $\permittivity$ the vacuum permittivity, $\charge_j$ the charge of the particle $j$, and $\relativePermittivity$ the relative permittivity of the material. Following the procedure described above, but now collecting the interaction-specific pre-factors, we find the respective coupling strengths:
\begin{equation}
    \begin{aligned}
    \couplingG = -\frac{\gravConst\mass}{2\freqRe\distanceD^3}, \qquad
    \couplingC = \frac{\charge_1\charge_2}{8\pi\permittivity\mass\freqRe\distanceD^3}, \qquad
    \couplingCasimir = -\frac{1449}{128}\frac{\hbar \speedoflight}{\pi^3}\frac{\mass}{\massdensity^2\freqRe\distanceD^9}\left(\frac{\relativePermittivity-1}{\relativePermittivity+2}\right)^2.
    \end{aligned} \label{eq:SM_gcouplings}
\end{equation}

\end{document}